\newcommand{\ie}{{\it i.e.}}
\newcommand{\eg}{{\it e.g.}}

\newcommand{\vs}{{\it vs.}}
\newcommand{\etc}{{\it etc.}}

\newcommand{\rhs}{r.h.s.}
\newcommand{\BR}[1]{\linebreak[0]#1\linebreak[0]}
\newcommand{\onlinecite}[1]{\cite{#1}}

\documentstyle[pre,aps]{revtex}

\draft

\begin{document}

\title{
  %{\bf PREPRINT v1.5 -- to be submitted to PRE}\\\ \\
  Josephson vortex in a ratchet potential: Theory
}

\author{
  E.~Goldobin\cite{gold-email}, 
}

\address{
  Institut f\"{u}r Schicht- und Ionentechnik (ISI),
  Forschungszentrum J\"ulich GmbH (FZJ),\\
  D-52425, J\"ulich, Germany
}

\author{A.~Sterck, and D.~Koelle}

\address{
  II. Physikalisches Intsitut,
  Universit\"{a}t zu K\"{o}ln,
  D-50937, K\"{o}ln, Germany
}

\date{\today}

%\wideabs{ %REVTeX 3.1

\maketitle

\begin{abstract}

We propose a new type of Josephson vortex ratchet. In this system a Josephson vortex moves in a periodic asymmetric potential under the action of a deterministic or random force with zero time average. For some implementations the amplitude of the potential can be controlled during the experiment, thus, allowing to tune the performance of the system and build rocking as well as flashing ratchets. We present a model describing the dynamics of the fluxon in such a system, show numerical simulation results, and discuss the differences between conventional and Josephson vortex ratchets. The investigation of this system may lead to the development of the fluxon rectifier --- a device which produces a quantized dc voltage from colored noise (non-equilibrium fluctuations).

\end{abstract}


%} %%End Of \WideAbs{...}

\section{Introduction}

To extract useful work from random motion was a dream of mankind since the days when the Brownian motion was recognized. Unfortunately, the second law of thermodynamics forbids to extract energy ``for free'' from equilibrium thermal fluctuations (white noise), which was didactically demonstrated by R.~Feynman in his {\it Lectures}\cite{Feynman:Lectures:Par46}. Nevertheless, one can extract useful work from {\em non-equilibrium} or {\em time-correlated} (colored) noise ``not paying'' for it, using so-called ratchets, \ie, systems with an asymmetric periodic potential\cite{Magnasco:Forced-Thermal-Ratchets}. Recently there was a boost of activity in this field related to the experimental investigation of directed motion in biological systems, so-called Brownian motors which, \eg, move muscles or transport vesicles in a cell\cite{Maddox:Nature3Views}. In the latter case the probable mechanism of operation is the motion of kinesin molecules along the surface of microtubules, which can be mapped to the motion of Brownian particles along a one-dimensional ratchet potential with the period $8.2\,{\rm nm}$\cite{Svoboda}. The non-equilibrium energy is supplied by chemical reaction of splitting of adenosine triphosphate (ATP) which takes place close to the kinesin molecule. 

In addition to the application of ratchets as noise rectifiers, it was suggested to use them for very efficient separation of small objects with different mobility, \eg, DNA molecules, viruses, \etc\cite{Leibler:RatSeparation,Kettner:DriftRat}. The particle separation is based on so-called deterministic ratchets\cite{Sarmiento:Det-Ratchet}, where the particles move in a certain direction under the action of a deterministic force with zero time average. Moreover, changing the force profile one can reverse the direction of the particle motion\cite{Sarmiento:Det-Ratchet}. The classification and discussion of different types of ratchet systems can be found in Ref.~\onlinecite{Haenggi:review}.

In this paper we focus on Josephson ratchets which are of particular interest because (a) the directed motion results in a dc voltage according to the Josephson relation and (b) these systems can operate at very high frequencies up to about $100\,{\rm GHz}$. As a first example we mention the asymmetric dc SQUID where the equation of motion for the Josephson phase (difference of quantum mechanical phases) corresponds to the motion of an imaginary particle in a 2D ratchet potential. Such SQUID ratchets have been proposed \cite{Zapata:Propose-SQUID-Rats} and studied experimentally\cite{Weiss:Exp-SQUID-Rats}. Another type of Josephson ratchet investigated recently is a 1D array of Josephson junctions with spatially modulated properties\cite{Jos-Arr-Rats}. A Josephson kink (vortex), which can move along the array, can be considered as quasi-particle in a 1D ratchet potential.

Here we propose a new class of Josephson ratchets which further develops the idea of a kink in a 1D array. The proposed system consists of a 1D long Josephson junction (LJJ) which may be bent in the $ab$-plane [see Fig.~\ref{Fig:circuit}(a)] or have variable width $w(x)$. Here and below $x$ is a curvelinear coordinate along the junction. The fluxon (Josephson vortex) moving along the junction, from a mathematical point of view, is a topological soliton. It has its own mass, velocity, and other particle-like properties\cite{FluxonProps}. We study the motion of a fluxon in LJJs in a ratchet potential which can be formed either by applying an external magnetic field and bending the junction properly or by modulating its width $w(x)$. To provide the required periodicity of the potential, the junction is topologically closed in a loop. Such a geometry is similar to the well known annular Josephson junction\cite{Davidson:AnnLJJ:Theor,Davidson:AnnLJJ:Exp,Ustinov:AnnLJJ,Martucello:ALJJ:Model,GrJensen:ALJJ-with-H} in which a fluxon moves in a sinusoidal potential created by a magnetic field. Using a more elaborated shape\cite{Ustinov:ShapedALJJ} one can form an asymmetric potential with the possibility to control its amplitude by changing the amplitude of the external magnetic field. An alternative idea of magnetic field modulation using a specially shaped control line is proposed in Ref.~\onlinecite{Carapella:Rat}. The directional motion of a fluxon can be detected by measuring the dc voltage across the junction which is, due to the Josephson relation, proportional to the average velocity of a fluxon.

Before discussing fluxon dynamics in a ratchet potential we would like to stress the difference between conventional ratchets and Josephson vortex ratchets. First, the fluxon, although it has a topological charge and in general behaves like a particle, is a nonlinear wave, \ie, it can change its shape rather strongly as well as emit electromagnetic waves. Second, the fluxon dynamics is usually studied in the underdamped limit which is opposite to the overdamped case which was considered for the majority of work on Brownian particles in a ratchet potential. Small damping may result in chaotic dynamics even in the deterministic case\cite{Mateos:Chaos-in-Det-Rat} and, in the case of the fluxon, even without any potential\cite{Goldobin:Chaos-Ann-ac-drive}. Thus, to have a well defined behavior of a fluxon ratchet, one has to work in the overdamped or in the weakly underdamped limit. If one uses conventional Nb-AlO$_x$-Nb technology to fabricate the LJJ, this requirement means that the working temperature should be very close to $T_c$. As an alternative one can use junctions with intrinsically high damping such as SINIS LJJs\cite{SINIS} or high-$T_c$ LJJs technology\cite{HTSJJ} which allows to fabricate LJJs of required topology. Third, if we consider multi-particle dynamics, the strong repelling interaction between fluxons plays an important role and must be taken into account.

This paper is organized as follows. In section \ref{Sec:Model} we derive the equations for the dynamics of the Josephson phase in a bent LJJ of variable width $w(x)$ in the external magnetic field. We also discuss different kinds of fluxon ratchets, their advantages and drawbacks. The numerical simulation results are presented in \ref{Sec:Simulation}. Section \ref{Sec:Conclusion} concludes this work.

\section{The Model}
\label{Sec:Model}

Here we derive the generalized perturbed sine-Gordon equation which takes into account the curvature of the LJJ in the $ab$-plane [see Fig.~\ref{Fig:circuit}(a)], the uniform magnetic field ${\bf \vec{H}}$ applied in the plane of the junction (in $b$ direction) as well as the modulation of the LJJ width $w(x)$ along its length $x$.

We start from a discrete representation of the LJJ shown schematically in Fig.~\ref{Fig:circuit}(b). The Kirchhoff equations for the Josephson phases in the cell and for the currents in one of the nodes are:
\begin{eqnarray}
  \phi(x+dx)-\phi(x) = \frac{2\pi}{\Phi_0}\Phi(x)
  = \frac{2\pi}{\Phi_0}\left[ \Phi_e(x) - L(x)I_L(x) \right]
  ; \label{Eq:Kirchhoff:phases}\\
  I_L(x) - I_L(x-dx) = I_e(x) - I(x)
  , \label{Eq:Kirchhoff:currents}
\end{eqnarray}
where $\phi(x)$ is the Josephson phase at point $x$ of the junction, $\Phi(x)$ and $\Phi_e(x)$ are the total and the external magnetic flux applied to the cell, respectively, $L(x)$ is the inductance of the piece of the junction electrodes between $x$ and $x+dx$, $I_L(x)$ is the current in the electrodes, \ie{} through the inductance $L(x)$, $I_e(x)$ is the externally applied bias current, and $I(x,t)$ is the current through the Josephson junction. The particular expression for $I(x,t)$ depends on the adopted JJ model and is introduced later.

Assuming that the interval $dx$ is infinitesimal, we can rewrite Eqs.~(\ref{Eq:Kirchhoff:phases}) and (\ref{Eq:Kirchhoff:currents}) in a differential form using the following expressions:
\begin{eqnarray}
  I(x)     &=& j(x) w(x) dx
  ; \label{Eq:I(x)-via-j(x)}\\
  I_e(x)   &=& j_e(x)w(x) dx
  ; \label{Eq:Ie(x)-via-je(x)}\\
  L(x)     &=& \frac{\mu_0 d'}{w(x)} dx
  ; \label{Eq:L(x)-via-l(x)}\\
  \Phi_e(x)&=& \mu_0( {\bf \vec{H}\cdot\vec{n}} ) \Lambda dx
  = \mu_0 H(x) \Lambda dx
  , \label{Eq:Phi_e}
\end{eqnarray}
where $\mu_0 d'$ is the inductance of one square of the superconducting electrodes\cite{Likharev:book}, $d'\approx2\lambda_L$ is the effective magnetic thickness of the junction\cite{Likharev:book}, ${\bf \vec{n}}$ is the unit vector normal to the plane of the junction cell as shown in Fig.~\ref{Fig:circuit}(b), $\Lambda\approx2\lambda_L$ is the effective penetration depth of the magnetic field into the junction\cite{Likharev:book}, and $\lambda_L$ is the London penetration depth of the superconducting electrode. We assume that the films are spatially uniform so that $d'$ and $\Lambda$ are independent on $x$.

Substituting Eqs.~(\ref{Eq:I(x)-via-j(x)})--(\ref{Eq:Phi_e}) into Eqs.~(\ref{Eq:Kirchhoff:phases}) and (\ref{Eq:Kirchhoff:currents}) we can rewrite the latter in a differential form as 
\begin{eqnarray}
  \frac{\partial\phi}{\partial x}
  = \frac{2\pi}{\Phi_0}
  \left[
    \mu_0 H(x) \Lambda
    - \frac{\mu_0 d'}{w(x)} I_L(x)
  \right]
  ; \label{Eq:Kirchhoff:phases:diff}\\
  \frac{\partial I_L(x)}{\partial x} = w(x) \left[ j_e(x) - j(x) \right]
  . \label{Eq:Kirchhoff:currents:diff}
\end{eqnarray}

Excluding $I_L(x)$ from Eqs.~(\ref{Eq:Kirchhoff:phases:diff}) and (\ref{Eq:Kirchhoff:currents:diff}), we get the equation which describes the dynamics of the Josephson phase in the system:
\begin{equation}
  w(x)\left[ j_e(x) - j(x) \right] =
  \frac{1}{\mu_0d'}\frac{d}{dx}\left\{ w(x)\left[
    \mu_0 \Lambda H(x) - \frac{\Phi_0}{2\pi} \phi_x
  \right]\right\}
  , \label{Eq:Final:Phys:NoModel}
\end{equation}
Here and below, the subscripts $t$ and $x$, if any, denote the derivatives with respect to time $t$ and coordinate $x$, respectively. Note, that we didn't include any particular model of JJ into our equation up to now, which is a definite advantage of this derivation procedure. In the case of the simple RSJ model, one should substitute $j(x)$ which is the sum of the supercurrent, normal (quasiparticle) current and displacement current densities:
\begin{equation}
  j(x) = j_c \sin(\phi) 
  + \frac{\Phi_0}{2\pi R} \phi_t 
  + C\frac{\Phi_0}{2\pi}\phi_{tt}
  \label{Eq:RSJ-model}
\end{equation}
into Eq.~(\ref{Eq:Final:Phys:NoModel}). Here $j_c$, $R$ and $C$ are the critical current density, specific resistance and specific capacitance of the junction, accordingly. In this case Eq.~(\ref{Eq:Final:Phys:NoModel}) can be rewritten in a form which resembles the usual sine-Gordon equation\cite{Likharev:book}:
\begin{equation}
  \lambda_J^2 \phi_{xx} - \omega_p^{-2}\phi_{tt} - \sin(\phi)
  =\omega_c^{-1}\phi_t - \gamma(x)
  + Q H_x(x)
  + \frac{w_x(x)}{w(x)} \left[ Q H(x) - \lambda_J^2 \phi_x \right]
  , \label{Eq:Final:Phys}
\end{equation}
where $\lambda_J=\sqrt{\Phi_0/(2 \pi \mu_0 j_c d')}$ is the Josephson penetration depth, $\omega_p=\sqrt{2 \pi j_c/(\Phi_0 C)}$ is the Josephson plasma frequency, $\omega_c=2 \pi j_c R/\Phi_0$ is the so-called critical frequency, $\gamma(x)=j_e(x)/j_c$ is a normalized bias current density, and $Q=2\pi\mu_0\Lambda\lambda_J^2/\Phi_0$.

For theoretical investigation of the system we introduce standard normalized units, \ie, we normalize the coordinate to the Josephson penetration depth $\lambda_J$, and the time to the inverse plasma frequency $\omega_p^{-1}$. After such simplifications, Eq.~(\ref{Eq:Final:Phys}) can be rewritten as\cite{Eqs:H(x),Eqs:w(x)}:
\begin{equation}
  \phi_{xx} - \phi_{tt} - \sin(\phi)
  =\alpha\phi_t - \gamma(x)
  + h_x(x) + \frac{w_x(x)}{w(x)} \left[ h(x) - \phi_x \right]
  , \label{Eq:Final:Norm}
\end{equation}
with the damping coefficient $\alpha=\omega_p/\omega_c \equiv 1/\sqrt{\beta_c}$, and the field $h$ normalized in a usual way as
\begin{equation}
  h(x) = \frac{2H(x)}{H_{c1}}
  . \label{Eq:FieldNormalization}
\end{equation}
$H_{c1}=\Phi_0/(\pi\mu_0\Lambda\lambda_J)$ is the first critical field (penetration field) for a LJJ which is, in fact, equal to the field in the center of the fluxon. The normalized velocity is given in natural units of ${\bar c}_{0}=\lambda_J\omega_p$, where ${\bar c}_{0}$ is the so-called Swihart velocity. The normalized voltage $V=\phi_t$ is given in units of $\Phi_0\omega_p/(2\pi)$. From now on all quantities are given in normalized units.

In comparison with the usual perturbed sine-Gordon equation, Eq.~(\ref{Eq:Final:Norm}) contains 3 additional terms. The term $h_x(x)$ describes the effect of the applied magnetic field when the junction is bent in the $ab$-plane. The second term $\left[ w_x(x)/w(x) \right]\phi_x$ comes from the width modulation. The last term $\left[ w_x(x)/w(x) \right]h(x)$ describes the mixture of both and does appear only when both, field modulation due to curvature and width modulation are present.

It can be checked by a direct substitution into the Euler-Lagrange equation
\begin{equation}
   \frac{d}{dt}\frac{\partial{\cal L}}{\partial\phi_t}
  +\frac{d}{dx}\frac{\partial{\cal L}}{\partial\phi_x}
  -\frac{\partial{\cal L}}{\partial \phi} = 0
  , \label{Eq:Lagrange}
\end{equation}
that the Lagrangian density
\begin{equation}
  {\cal L} = w(x)
  \left\{
    \frac{\phi_t^2}{2} - \frac{\left[ \phi_x - h(x)\right]^2}{2}
    -\left( 1 - \cos\phi \right)
  \right\}
  , \label{Eq:LagrangianDensity}
\end{equation}
results in the equation of motion (\ref{Eq:Final:Norm}) without the $\alpha\phi_t$ and $\gamma$ terms which describe dissipation and external force and therefore are not included in the Lagrangian density. Of course, the Lagrangian density (\ref{Eq:LagrangianDensity}) can be obtained directly from Fig.~\ref{Fig:circuit}(b) and the RSJ model. From Eq.~(\ref{Eq:LagrangianDensity}) one can see that instead of the usual potential energy term $\phi_x^2/2$ we now have $w(x)\left[ \phi_x - h(x) \right]^2/2$, \ie, actually 3 terms. The first, $w(x)\phi_x^2/2$, is the obvious generalization of the usual potential energy term to the case of variable width $w(x)$. The second term $w(x)h(x)^2/2$ is a constant term due to the applied field and {\em is not} related to the fluxon motion or other Josephson phase activity in the junction. In fact, there are no traces of this term in Eq.~(\ref{Eq:Final:Norm}). The third term $h(x)\phi_x w(x)$ represents the part of the potential energy density which we are interested in and which we are going to exploit to build a system with a ratchet potential.

One of the solutions of the sine-Gordon Eq.~(\ref{Eq:Final:Norm}) with zero \rhs{} is a soliton (fluxon)
\begin{equation}
  \phi(x,x_0) = 4 \arctan\exp(x-x_0)
  , \label{Eq:soliton}
\end{equation}
with the center situated at point $x_0$. We consider non-relativistic motion \ie, $dx_0/dt \ll 1$. Further assuming that the fluxon profile (\ref{Eq:soliton}) does not change much due to the \rhs{} of Eq.~(\ref{Eq:Final:Norm}) which acts as a perturbation, we can get the explicit expression for the potential energy $U(x_0)$ as a function of fluxon coordinate $x_0$. For this purpose, we note that in expression (\ref{Eq:LagrangianDensity}) for the Lagrangian density the second and the third terms correspond to the potential energy density ${\cal U}(x,x_0)$ (with opposite sign). The potential energy $U(x_0)$ is obtained by substituting $\phi(x,x_0)$ from Eq.~(\ref{Eq:soliton}) into ${\cal U}(x,x_0)$ and integrating over $x$. Thus, we get
\begin{equation}
  U(x_0) 
  = \int_{-\infty}^{+\infty}
    \frac{4w(x)}{\cosh^2(x-x_0)}
  - \frac{2w(x)h(x)}{\cosh(x-x_0)}
  \:dx 
  . \label{Eq:U(x)}
\end{equation}
The first term corresponds to the potential energy due to width modulation, the second due to shape, field and width. 

The first possibility to form a ratchet potential is to apply no magnetic field ($h=0$) and to vary the width $w(x)$ of the junction. In this case the potential energy will be given by the first term of Eq.~(\ref{Eq:U(x)}). Moreover, when the junction width $w(x)$ does not change much over the distance comparable with the fluxon size, the potential
\begin{equation}
  U(x_0) 
  \approx 8w(x_0)
  \label{Eq:U(x_0):h=0&w-slow}
\end{equation}
just repeats the $w(x_0)$ profile. From Eq.~(\ref{Eq:U(x_0):h=0&w-slow}) it follows that if the width changes as a saw-tooth, so does the potential energy, except for the vicinity of the saw-tooth's infinite slope where one has to calculate the convolution according to Eq.~(\ref{Eq:U(x)}). Here and below when mentioning saw-tooth profile we mean a saw-tooth with finite positive slope and infinite negative slope. In the case of the saw-tooth where the width changes from $w_0$ to $w_0+\Delta w$ at the point $x_0=0$ (infinite slope), the potential energy in the vicinity of the point $x_0=0$ is $U(x_0)=8w_0+4\Delta w \left[ 1+\tanh(x_0) \right]$. The practical implementation of such a ratchet would look like an annular LJJ with the outer edge having the shape of a circle and inner edge having the shape of one turn of a spiral. This geometry has the advantage that the corresponding potential can be made ideally saw-tooth-like except for the smearing due to the convolution in Eq.~(\ref{Eq:U(x)}), which is a common feature of all fluxon based systems. In some sense, this system is an analog of the Josephson ratchets based on 1D arrays\cite{Jos-Arr-Rats}. Unfortunately both types of ratchets do not allow to control the potential height during experiment which can be considered as a disadvantage.

The second possibility is to keep $w$ constant, to apply a magnetic field and to bend the junction in the $ab$-plane. In this case the first term of Eq.~(\ref{Eq:U(x)}) gives a constant and we have to consider only the second term. Again, if $h(x)$ changes slowly in comparison with the fluxon size, Eq.~(\ref{Eq:U(x)}) is simplified to
\begin{equation}
  U(x_0) 
  = - 2 \pi w h(x_0)
  . \label{Eq:U(x):w=const:h-slow}
\end{equation}
In the well known case of the ring shaped junction, the field $h(x)=h_0\cos(\theta)=h_0\cos(x/R)$ [$\theta$ is the angle between ${\bf \vec{H}}$ and ${\bf \vec{n}}$, as shown in Fig.~\ref{Fig:circuit}], and therefore we get a symmetric potential. If we deform the ring properly, the potential can be made asymmetric as desired. Possible experimental shapes are shown in Fig.~\ref{Fig:shapes}. The advantage of this kind of ratchets is that one can control the amplitude of the potential during experiment by varying the amplitude $h_0$ of the magnetic field. The possibility to tune the potential height allows to implement so-called flashing ratchets\cite{Haenggi:review} by applying either an ac magnetic field using a coil or just placing an rf antenna close to the junction so that the $\vec{\bf H}$ of the emitted electromagnetic wave will have non-zero component in the $ab$-plane.

Note, that the terms $\gamma(x)$ and $h_x(x)$ in Eq.~(\ref{Eq:Final:Norm}) from a mathematical point of view play the same role. Therefore in the experiment the field $h(x)$ can be substituted by a properly chosen additional bias current $\gamma_p(x)=-h_x(x)$, which has zero average in space. The inverse mapping is also valid, but the bias current with non-zero average maps to a non-periodic field (potential) with linearly growing component, which does not belong to the class of ratchets. In section \ref{Sec:Simulation}.A we, in fact, use $\langle\gamma\rangle \ne 0$ (these brackets denote spacial averaging), but only to test the asymmetry of the potential. When we demonstrate the real operation of the ratchet in the section \ref{Sec:Simulation}.B, we have $\langle\gamma\rangle = 0$.

\section{Simulation results}
\label{Sec:Simulation}

In this section we study the fluxon dynamics in a Josephson ratchet of the second type ($h \ne 0$, $w={\rm const}$) for a saw-tooth field profile $h(x)$. This ideal asymmetric profile is not only of academic interest because it should show the best figures of merit for fluxon ratchets but also can be quite closely reproduced in a real system [see shape in Fig.~\ref{Fig:shapes}(b) and corresponding $h(x)$ in Fig.~\ref{Fig:Explanation}(a)]. All simulations were performed using an explicit numerical scheme for Eq.~(\ref{Eq:Final:Norm}) using a LJJ of the normalized length $\ell=20$, with damping coefficient $\alpha=0.2$ (weakly underdamped limit). The numerical technique and simulation software are discussed in detail elsewhere\cite{StkJJ}.

\subsection{Probing the asymmetry of the potential}
\label{Sec:Sim:Test}

First, we probe the asymmetry of the potential by calculating critical current $\gamma_c$ \vs{} potential height $h_0$ for the case of one trapped fluxon [$\gamma_{c1}^\pm(h_0)$] and for the case of no trapped fluxons [$\gamma_{c0}^\pm(h_0)$]. The superscripts `$+$' or `$-$' correspond to opposite directions of applied bias current. 

On the basis of the model derived in the previous section we can understand how these dependences should look like for arbitrary magnetic field profile $h(x)$. In general, due to the loop-like geometry, the left and right tails of a fluxon can interact. We assume that the junction is long enough and this interaction in negligible. This situation is equivalent to the long periodic system where the fluxons are separated by a large distance $\ell\gg\lambda_J$. Furthermore, we represent the field $h(x)$ as $h_0{\cal H}(x)$ and map it to the equivalent additional bias current $\gamma_p(x)=-h_0{\cal H}_x(x)$. An example of $h(x)$ and $\gamma_p(x)$ derived from the geometry in Fig.~\ref{Fig:shapes}(b) is shown in Fig.~\ref{Fig:Explanation}(a).

If there is a fluxon in the junction, $\gamma(x)$ and $\gamma_p(x)$ translate into a driving force $F_\gamma(x_0)$ and a potential (pinning) force $F_p(x_0)$, respectively, acting on the fluxon\cite{McLoughlinScott}
\begin{equation}
  F_\gamma(x_0)=\int_{-\infty}^{+\infty} 
  \frac{2\gamma(x)w}{\cosh(x-x_0)} \:dx 
  = 2 \pi \gamma w
  , \label{Eq:Fgamma}
\end{equation}
\begin{equation}
  F_p(x_0)=\int_{-\infty}^{+\infty} 
  \frac{2\gamma_p(x)w}{\cosh(x-x_0)}\:dx 
  = h_0 f(x_0) w
  . \label{Eq:Fp}
\end{equation}
The corresponding potential $U(x_0)$ and force $f(x_0)$ are shown in Fig.~\ref{Fig:Explanation}(b). For the sake of simplicity we suppose that $\gamma$ does not depend on $x$, but our discussion can be easily generalized to the case when the \rhs{} of Eq.~(\ref{Eq:Fgamma}) is equal to $2 \pi \Gamma(x_0)w$. The pinning force (\ref{Eq:Fp}) can be also obtained directly from expression (\ref{Eq:U(x)}) for the potential energy. When we increase $\gamma$ the fluxon is pinned while these two forces can compensate each other, \ie, $F_\gamma(x_0)+F_p(x_0)=0$. The depinning happens for 
\begin{equation}
  \gamma=\gamma_{c1}=-h_0 f(x_1)/2\pi
  , \label{Eq:Gamma-depinning}
\end{equation}
where $x_1$ is the coordinate at which $f(x_1)$ has a minimum [see Fig.~\ref{Fig:Explanation}(b)]. We assume that $\gamma>0$ and $f(x_1)<0$. Thus from Eq.~(\ref{Eq:Gamma-depinning}) we see that $\gamma_{c1}(h_0)$ looks like a straight line which starts from the origin.

On the other hand, regardless of the presence of a fluxon, the Josephson phase evolves under the action of the net current $\gamma+\gamma_p$. From Fig.~\ref{Fig:Explanation}(a) it is clear that at point $x_2$, where $\gamma_p(x)$ is maximum, this sum can exceed 1 for some value of $\gamma=\gamma_c$ and the junction switches to the resistive state. Using the $\gamma_p=-h_0{\cal H}_x(x)$ the condition $\gamma_c+\gamma_p(x_2)=1$ gives 
\begin{equation}
  \gamma_c(h_0)
  =1+h_0{\cal H}_x(x_2)
  =1-h_0\left| {\cal H}_x(x_2) \right|
  , \label{Eq:Gamma-c}
\end{equation}
\ie, also a straight line, but with the negative slope.

Summarizing, $\gamma_{c1}(h_0)$ has two branches (\ref{Eq:Gamma-depinning}) and (\ref{Eq:Gamma-c}), and follows the one with the lowest critical current for given $h_0$. For fields $0<h_0<h_*$, $\gamma_{c1}(h_0)$ is given by Eq.~(\ref{Eq:Gamma-depinning}), and for $h_0>h_*$ by Eq.~(\ref{Eq:Gamma-c}), where the field
\begin{equation}
  h_* = \frac{-1}{f(x_1)/2\pi+{\cal H}_x(x_2)}
  , \label{Eq:h*}
\end{equation}
is the field where these two dependences intersect. The dependence $\gamma_{c0}(h_0)$ has only one branch (\ref{Eq:Gamma-c}) since the fluxon depinning mechanism is absent.

When one applies the current $\gamma$ in the opposite direction, the dependences qualitatively look the same but the particular values of slopes {\em can} be different in the case of an asymmetric potential, if $\gamma_p(x_2)\ne-\gamma_p(x_4)$ and $f(x_1)\ne-f(x_3)$ (see Fig.~\ref{Fig:Explanation}). Thus, the measurement of $\gamma_{c0}^\pm(h_0)$ and $\gamma_{c1}^\pm(h_0)$ gives direct information about the asymmetry of the field and of the potential, respectively. Note that an asymmetric potential {\em may} have $\gamma_p(x_2)=-\gamma_p(x_4)$ and/or $f(x_1)=-f(x_3)$, \ie, it does not reveal its asymmetry in the measurements of $\gamma_{c0}^\pm(h_0)$ and $\gamma_{c1}^\pm(h_0)$. Inversely, the asymmetry of $\gamma_{c0}^\pm(h_0)$ and $\gamma_{c1}^\pm(h_0)$ is a clear indication of asymmetry in the system.

We stress that the first branch of $\gamma_{c0}^\pm(h_0)$ and the second branch of $\gamma_{c1}^\pm(h_0)$ show the asymmetry of the field $h(x)$, while the first branch of $\gamma_{c1}^\pm(h_0)$ shows the asymmetry of potential $U(x_0)$, \ie, already after convolution (\ref{Eq:U(x)}) with $\phi_x$. While $h_x(x_0)$ may have appreciable asymmetry, $U(x_0)$ may be almost symmetric. 

We should note that the above analysis is valid for an arbitrary field profile $h(x)$ as long as $h_x(x)$ can be considered as a perturbation in the \rhs{} of Eq.~(\ref{Eq:Final:Norm}). However, for the case of the saw-tooth potential this is not the case, since $h_x(x)$ has a $\delta(x)$-like behavior at the point of the junction where the saw-tooth has infinite slope. Therefore our analysis is valid only qualitatively. In particular the dependences $\gamma_c^\pm(h_0)$ still have linear slopes but the values of the slopes can not be calculated using simple integrals as shown above because the fluxon shape will differ from (\ref{Eq:soliton}) quite considerably, as we saw in numerical simulations. Therefore we present the results of direct numerical simulation of Eq.~(\ref{Eq:Final:Norm}).

The simulated $\gamma_{c0}^\pm(h_0)$ and $\gamma_{c1}^\pm(h_0)$ for an ideal saw-tooth field profile $h(x)$ are presented in Fig.~\ref{Fig:Sim:Ic(H):saw}. The magnetic field is given in terms of the force amplitude $h_0 2 \pi/\ell$. If the system is closed in a loop with a normalized circumference $\ell$, the $h(x)$ can be written in the form $h_0{\cal H}(2\pi x/\ell)$ where ${\cal H}$ is geometry dependent $2\pi$ periodic function normalized to 1. Therefore the term $h_x(x)$ in Eq.~(\ref{Eq:Final:Norm}) is
\begin{equation}
  h_x(x) = h_0 \frac{2 \pi}{\ell} {\cal H}'\left( \frac{2 \pi}{\ell}x \right)
  , \label{Eq:h_x(x)}
\end{equation}
where ${\cal H}'(\xi)=d{\cal H}(\xi)/d\xi$ is a periodic function. This means that the potential force scales inversely proportional to the length of the junction $\ell$. So, to get rid of the length dependence and make our results valid for any length $\ell \gg 1$, we present our results as a function of $h_0 2 \pi /\ell$.

From Fig.~\ref{Fig:Sim:Ic(H):saw} one can see that the depinning current $\gamma_{c1}(0)=0$ and then grows linearly with field, as expected from the theory, up to some field $h_*2\pi/\ell\approx1.8$. After that it decreases linearly with field, also according to our prediction, and, finally, exhibits multiple branches corresponding to the multi-fluxon states.

As it should be in the case of a saw-tooth potential, the slopes of $\gamma_{c0}^\pm(h_0)$ and $\gamma_{c1}^\pm(h_0)$ are not symmetric, and the ratio of the slopes is about $4$. Ideally, if the fluxon were a real particle we could expect an infinite force necessary to push the fluxon out of the well in the direction against the infinite slope and, therefore, an infinite ratio of slopes on $\gamma_{c1}^\pm(h_0)$ curves. In practice, due to the convolution (\ref{Eq:U(x)}), we have a finite force and a finite ratio of slopes on the $\gamma_{c1}^\pm(h_0)$ dependences. The finite ratio of slopes on $\gamma_{c0}^\pm(h_0)$ as well as of the second slopes $\gamma_{c1}^\pm(h_0>h_*)$ is observed because perturbation theory does not apply as mentioned above.

The next step is the understanding of the {\em dynamics} of the fluxon in an asymmetric potential. One of the simplest observations which can be made numerically as well as experimentally is the examination of the  fluxon trapping current $\gamma_{\rm tr}$, \ie, the minimum current at which a fluxon still moves along the system, not being trapped by the potential. Obviously, in the underdamped system the trapping of the fluxon by the potential will not take place while the kinetic energy of the fluxon exceeds the height of the potential, and one should not see any difference in the fluxon trapping currents $\gamma_{\rm tr}^\pm(h_0)$ for opposite bias current directions. In the strongly overdamped case the fluxon dissipates energy so fast that to move further under the action of the driving force, the driving force should always overcome the maximum value of the potential (trapping) force. In an asymmetric potential the maximum force created by the potential is different for opposite directions of fluxon motion and we should expect a difference in the fluxon trapping current. The simulated $\gamma_{\rm tr}^\pm(h_0)$ curves for a saw-tooth potential are shown in Fig.~\ref{Fig:Sim:Itr(H):saw}. Even in the slightly underdamped case $\alpha=0.2$ the dependences for positive and negative direction of fluxon motion (bias current) differ quite considerably. We also note that $\gamma_{\rm tr}^+(h_0)$ almost coincides with $\gamma_{c1}^+(h_0)$  (see Fig.~\ref{Fig:Sim:Ic(H):saw}) for $h_0<2$ which means that for positive direction of bias there is no hysteresis on the $I$--$V$ characteristic when the system switches from zero voltage state to the state with moving fluxon and back. Instead, for the negative bias the hysteresis is present as can be seen in Fig.~\ref{Fig:IVC@H=1.0}.

\subsection{Motion due to monochromatic force}
\label{Sec:Sim:monochrom-drive}

In the following we investigate the motion of a fluxon in a system with the saw-tooth field profile $h(x)$ and $w={\rm const}$ under the action of a monochromatic ac bias current $\gamma=\gamma_{\rm ac}\sin(\omega t)$ in Eq.~(\ref{Eq:Final:Norm}). If as a result of an ac drive the fluxon starts to move around the junction, we can estimate the characteristic frequency of this process. It is clear that the maximum velocity of the fluxon is equal to the Swihart velocity (1 in normalized units) therefore the maximum revolution frequency is $\omega_{\max}=2\pi/\ell\approx0.314$.

In the limit of low frequency (quasi-static) drive $\omega\ll\omega_{\max}$ we can calculate the net average velocity of the fluxon, just integrating the current-velocity characteristics $u(\gamma)$ (see Fig.~\ref{Fig:IVC@H=1.0}) at a given value of field $h_0$ as: 
\begin{equation}
  u(\gamma_{\rm ac})
  =\frac{1}{2\pi}\int_0^{2\pi}u(\gamma_{\rm ac}\sin(\tau))\:d\tau
  , \label{Eq:u(gamma_ac)}
\end{equation}
where the integration path should follow the proper hysteretic branches of the $u(\gamma)$ dependence. 

In the case when $\omega$ is comparable with the maximum frequency of the fluxon rotation around the junction $\omega_{\max}$, we perform direct numerical simulations. The dependences $u(\gamma_{\rm ac})$ calculated numerically for different values of $\omega$ are shown in Fig.~\ref{Fig:Sim:Vdc(Irf):saw-tooth}(a)--(c) for $h_0 2\pi/\ell=0.5$, 1.0 and 2.0, respectively. The upper curves marked as ``qs'' (quasi-static) are calculated using Eq.~(\ref{Eq:u(gamma_ac)}). All other curves for $\omega=0.01$--0.1 were obtained by means of direct numerical simulation of Eq.~(\ref{Eq:Final:Norm}). Such values of frequencies were chosen taking into account that they should be less but comparable with $\omega_{\max}$. As we saw in simulations, at $\omega\ge0.2$ [in the case of Fig.~\ref{Fig:Sim:Vdc(Irf):saw-tooth}(a) already for $\omega=0.1$] the rectification is suppressed almost completely.

From Fig.~\ref{Fig:Sim:Vdc(Irf):saw-tooth} one can see that the dependence $u(\gamma_{\rm ac})$ for quasi-static case is piecewise with jumps at the values of bias current where switching between different hysteretic branches occurs. For example, let us focus on $h_02\pi/\ell=1.0$ [see Fig.~\ref{Fig:Sim:Vdc(Irf):saw-tooth}(b)]. In this case the switch between different hysteretic branches occurs at $\gamma$ which is equal to 0.522, 0.696, 0.916 (c.f. Figs.~\ref{Fig:Sim:Ic(H):saw}, \ref{Fig:IVC@H=1.0} and \ref{Fig:Sim:Vdc(Irf):saw-tooth}). Several characteristic regions can be distinguished. First, in the region of small values of $\gamma_{\rm ac}$, where $u=0$, the amplitude of the ac current is not big enough to exceed the pinning current and the fluxon is localized in the well. At higher ac bias up to $\gamma_{\rm ac}=0.522$ the positive depinning current is exceeded and the system switches to the fluxon step, \ie, the fluxon escapes from the well and starts to move around the junction. Note that during a period of ac drive the fluxon can only make integer number of turns and at the end of the period will be again localized in the well. This results in the quantization of the average velocity $u$, as can be seen as steps on all curves except for the quasi-static one. The step number corresponds to the number of turns done by the fluxon during one period of ac drive. The voltage rectifier based on this principle will give a quantized voltage  $V_n=n\Phi_0\omega/2\pi$ (in physical units) which is defined by fundamental constants and the applied frequency. The corresponding velocity quanta in Fig.~\ref{Fig:Sim:Vdc(Irf):saw-tooth} are given by $u_n=n\omega\ell/(2\pi)$. This gives $u_1\approx0.0318$ for $\ell=20$ and $\omega=0.01$. Note that the voltage quantum increases with $\omega$ and for $\omega=0.01$ five quanta (turns per ac cycle) give the same voltage as one quantum for $\omega=0.05$. For the quasi-static case the voltage quantum is infinitesimal so that the quasi-static curves look smooth. 

A further increase of the ac bias amplitude $0.522<\gamma_{\rm ac}<0.696$ results in a decrease of the dc voltage because during the negative semi-period of the ac drive the fluxon starts moving in the opposite direction so that the efficiency of the fluxon ratchet drops.

The big peak at $0.696<\gamma_{\rm ac}<0.916$ in Fig.~\ref{Fig:Sim:Vdc(Irf):saw-tooth}(b) is related to the switching of the junction to the resistive state during the positive semi-period while during the negative semi-period the junction stays in the zero-voltage state or on the fluxon step. When the amplitude of ac drive gets big enough to switch the system to the resistive state also during negative semi-periods, the dc voltage drops again and in the limit of very strong ac drive approaches zero. A large amplitude of ac bias $\gamma_{\rm ac} \gg 1$ implies that the system spends almost the whole period in the positive or negative resistive state and only a tiny fraction of the period on the asymmetric part of the current-velocity characteristic at low currents, so that the resulting average velocity is close to zero. In addition, such a strong ac drive results in chaotic dynamics, as can be seen in Fig.~\ref{Fig:Sim:Vdc(Irf):saw-tooth} for large values of $\gamma_{\rm ac}$.

Figure~\ref{Fig:Sim:Vdc(Irf):saw-tooth} also shows that the rectification effect decreases when the driving frequency increases and approaches the maximum fluxon rotation frequency $\omega_{\max}$.

We point out that the performance of the fluxon rectifier depends on the chosen potential amplitude $h_0$. For comparison in Fig.~\ref{Fig:Sim:Vdc(Irf):saw-tooth}(a)--(c) we show the rectification characteristics for different values of magnetic field amplitude $2\pi h_0/\ell=0.5$, 1.0 and 2.0 (see Fig.~\ref{Fig:Sim:Ic(H):saw}). We see that the potential depth affects the amplitude of ac drive at which rectification appears, at which $u(\gamma_{\rm ac})$ has maximum and it affects the relative location of the two regions corresponding to the fluxon rectification regime and resistive rectification regime. We note, that the resistive rectification regime in principle gives a larger effect which is inversely proportional to the damping coefficient $\alpha$, \ie, it can be made rather big. However, this is a rather trivial effect which can be obtained using any non-symmetric $I$--$V$ curve. The fluxon rectification regime, which is the main subject of this study, gives smaller but a quantized voltages. This can be an advantage for some applications.

\section{Conclusion}
\label{Sec:Conclusion}

We proposed a new type of Josephson vortex ratchets, where the motion of a fluxon along a long Josephson junction closed in a loop can be considered as the motion of a quasi-particle in a ratchet potential. The derived model suggests several implementations of the fluxon ratchets, and, in particular, the one where the amplitude of the potential can be controlled during experiment. Since the fluxon is a soliton (non-linear wave) moving in the underdamped medium with asymmetric potential, we expect some non-trivial effects related to (a) wave properties of the fluxon, (b) underdamped dynamics and (c) interaction between fluxons. As a first step, we performed a numerical analysis of the effective potential seen by a fluxon, and checked the rectification of a monochromatic signal. We found that a fluxon rectifier produces a quantized voltage, with a quantum given (due to the Josephson relation) only by the fundamental constant $\Phi_0$ and the driving frequency $\omega$. The experimental investigation of the proposed system based on (Nb-Al-AlO$_x$)-Nb Josephson junction technology is in progress.

\acknowledgments

We thank P.~Caputo, A.~Ustinov, M.~Fistul and B.~Malomed for fruitful discussions.

%%%%%%%%%%%%%%%%%%%%%

%%%%%%%%%%%%%%%%%%%%%%%%%%%%%%%%%%%%%%%%%%%%%%%%%%%%%%%%%%%

%
\begin{figure}
  \caption{
    Piece of LJJ ($d_1$ and $d_2$ are the thicknesses of the superconducting electrodes and $d_I$ is the thickness of the insulating tunnel barrier) which is bent in the $ab$-plane and has variable width $w(x)$: 
    (a) 3D view of geometry, 
    (b) schematic representation using discrete elements.
  }
  \label{Fig:circuit}
\end{figure}
\begin{figure}
  \caption{
    Possible shapes of LJJs (top view) that provide a ratchet potential when the magnetic field ${\bf \vec{H}}$ is applied in the direction shown by the arrow.
  }
  \label{Fig:shapes}
\end{figure}
\begin{figure}
  \caption{
    Illustration to the explanation of the $\gamma_{c0}^\pm(h_0)$ and $\gamma_{c1}^\pm(h_0)$ dependences. 
    (a) $h(x)$ and $\gamma_p(x)$. 
    (b) $U(x_0)$ and $f(x_0)$.
    These curves correspond to the LJJ shape shown in Fig.~\protect\ref{Fig:shapes}(b).
  }
  \label{Fig:Explanation}
\end{figure}
\begin{figure}
  \caption{
    Normalized critical currents $\gamma_{c0}^\pm(h_0)$ (no trapped fluxons) and $\gamma_{c1}^\pm(h_0)$ (one trapped fluxon) \vs{} magnetic field amplitude $h_0$ for the saw-tooth magnetic field profile $h(x)$ in the annular LJJ of length $\ell=20$. Dotted lines show the values of magnetic field at which rectification shown in Fig.~\protect\ref{Fig:Sim:Vdc(Irf):saw-tooth}(a)--(c) was calculated.
  }
  \label{Fig:Sim:Ic(H):saw}
\end{figure}
\begin{figure}
  \caption{
    Normalized fluxon trapping current $\gamma_{\rm tr}(h_0)$ \vs\ magnetic field amplitude $h_0$ for the saw-tooth magnetic field $h(x)$ in the annular LJJ of $\ell=20$ and $\alpha=0.2$. Dotted lines show the values of magnetic field at which rectification shown in Fig.~\protect\ref{Fig:Sim:Vdc(Irf):saw-tooth}(a)--(c) was calculated.
  }
  \label{Fig:Sim:Itr(H):saw}
\end{figure}
\begin{figure}
  \caption{
    Current-voltage characteristic of LJJ with one trapped fluxon shown as the dependence of fluxon velocity $u=V_{\rm dc}\ell/2\pi$ on the amplitude of the bias current $\gamma$ obtained for $\alpha=0.2$.
  }
  \label{Fig:IVC@H=1.0}
\end{figure}
\begin{figure}
  \caption{
    The dependence of the dc voltage across the junction $V_{\rm dc}$ given in terms of average fluxon velocity $u=V_{\rm dc}\ell/2\pi$ on the amplitude of the ac drive $\gamma_{\rm ac}$ for different normalized frequencies $\omega={\rm qs},\ 0.01,\ldots,0.1$ of the ac drive and $\alpha=0.2$. The curves corresponding to the different frequencies are intentionally shifted by $0.0636$ (two velocity quanta at $\omega=0.01$) relative to each other for the sake of better visibility. The insets of (a) and (b) show magnified views of the fluxon rectification region. Figs.~(a), (b) and (c) were obtained for $2\pi h_0/\ell=0.5$, 1.0 and 2.0, respectively.
  }
  \label{Fig:Sim:Vdc(Irf):saw-tooth}
\end{figure}


\begin{references}
\bibitem{FluxonProps}
  The inertial mass of a fluxon can be calculated as
  \[
    m = 4 \left( \frac{\Phi_0}{2 \pi} \right)
    \frac{j_c w}{\lambda_J\omega_p^2}
  \]
  Notations are explained in section \ref{Sec:Model}. For typical LJJ parameters the mass is $\sim 10^{-3}m_e$, where $m_e$ is the mass of the electron.

  The fluxon velocity can have the values from zero up to the so-called Swihart velocity ${\bar c}_{0}=\lambda_J\omega_p$ (see section \protect\ref{Sec:Model}) which is usually 20--40 times smaller than the speed of light.
\bibitem{Eqs:H(x)}
  An equation similar to (\ref{Eq:Final:Norm}) was used for the fist time in Ref.~\onlinecite{GrJensen:ALJJ-with-H} for the case $w={\rm const}$. The corresponding equation for an annular geometry and $w={\rm const}$ was derived in Ref.~\onlinecite{Martucello:ALJJ:Model}.
\end{references}
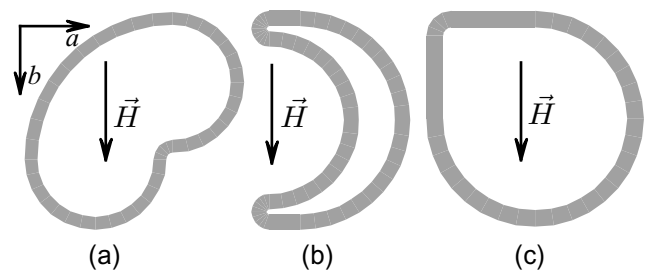
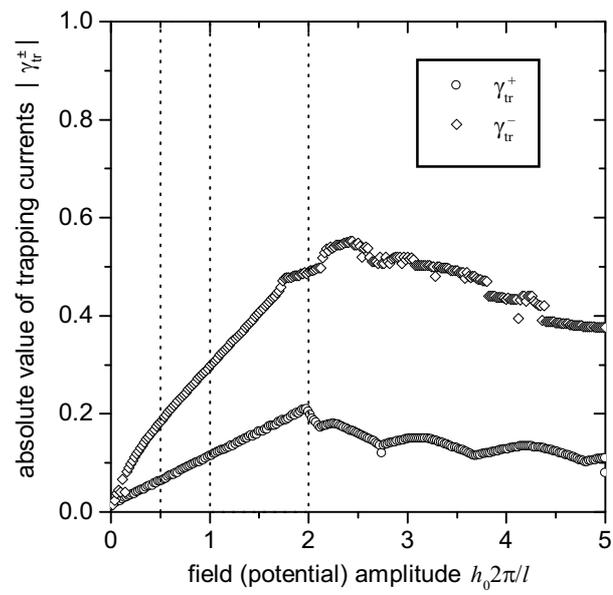
\end{document}